\documentclass[10pt,twoside]{article}
\usepackage{graphicx}
\usepackage{amsmath}
\usepackage{Latex-document}

\newtheorem{proposition}{Proposition}
\newtheorem{lemma}{Lemma}
\newcommand{\R}{\ifmmode{I\hskip -4pt R}
\else{\hbox{$I\hskip -4pt R$}}\fi}
\newcommand{\N}{\ifmmode{I\hskip -4pt N}
\else{\hbox{$I\hskip -4pt N$}}\fi}

\newcommand{\simleq}{\mathop{<}_{\sim} \limits}
\newcommand{\simgeq}{\mathop{>}_{\sim} \limits}
\newcommand{\ignore}[1]{}

\renewcommand{\thesection}{\arabic{section}}
\renewcommand{\thesubsection}{\thesection.\arabic{subsection}.}
\def\addsec{\addtocounter{section}{1}}

\title{\bf Cross-over in Scaling Laws:\vskip -2mm A Simple Example from Micromagnetics\vskip 6mm}
\author{Felix Otto\thanks{Department of Applied Mathematics, University of Bonn, Germany. E-mail: otto@iam.uni-bonn.de}
\vspace*{-0.5cm}}
\date{}

\begin{document}

\maketitle

\thispagestyle{first} \setcounter{page}{829}

\begin{abstract}

\vskip 3mm

Scaling laws for characteristic length scales (in time or in the model parameters) are both experimentally robust
and accessible for rigorous analysis. In multiscale situations cross--overs between different scaling laws are
observed. We give a simple example from micromagnetics. In soft ferromagnetic films, the geometric character of a
wall separating two magnetic domains depends on the film thickness. We identify this transition from a N\'eel wall
to an Asymmetric Bloch wall by rigorously establishing a cross--over in the specific wall energy.

\end{abstract}

\vskip 12mm

\section*{1. Introduction}\setzero \addsec

\vskip-5mm \hspace{5mm}

Many continuum systems in materials science display pattern
formation. These patterns are characterized by one or several
length scales. The scaling of these characteristic lengths in the
material parameters and/or in time are usually an experimentally
robust feature. These scaling laws, and their characterizing
exponents, are of interest to theoretical physics since they
express a certain universality. At the same time, scaling laws
(rather than more detailed features) are ameanable to heuristic
and rigorous analysis and thus are a good test for the model and a
challenge for mathematics.

Scaling laws and their exponents reflect a scale invariance. In a
multiscale model, these scale invariances are broken and only
approximately valid in certain parameter and/or time regimes. The
cross-over between two scaling laws reflects a change in the
dominant physical mechanisms. In studying cross-overs, theoretical
analysis may have an advantage over numerical simulation which has
to explore many parameter decades and thus has to cope with widely
separated length scales.

Together with various collaborators, the author has analyzed
scaling laws and their cross-overs in both static (variational)
and dynamic models. The dynamic models considered were of
gradient-flow type and thus endowed with a variational
interpretation: steepest descent in a multiscale energy landscape.
The examples are
\begin{itemize}
\item The branching of domains in uniaxial ferromagnets \cite{br}
(with R.\ Choksi and R.\ V.\ Kohn). Strongly uniaxial ferromagnets
have only two favored magnetization directions (``up'' and
``down''). The width of the corresponding domains decreases
towards a sample surface perpendicular to the favored axis. We
rigorously establish the scaling of the energy in the sample
dimensions in support of this behavior. To leading order, the
micromagnetic model behaves like a three-dimensional analogue of
the Kohn-M\"uller \cite{KM} model for twin branching.
\item The period of cross-tie walls in ferromagnetic films \cite{cr}
(with A.\ DeSimone, R.\ V.\ Kohn and S.\ M\"uller). Cross-tie
walls are transition layers between domains in ferromagnetic
films. They display a periodic structure in the tangential
direction. The experimentally observed scaling of the period in
the material parameters is not well-understood
\cite{HubertSchaefer}. In this paper, we present a combination of
heuristic and rigorous analysis which reproduces the experimental
scaling and thus identifies the relevant mechanism.
\item The rate of capillarity-driven spreading of a thin droplet \cite{lb}
(with L.\ Giacomelli). Here, the starting point is the lubrication
approximation. The scale invariant version of the model is
ill-posed and has to be regularized near the contact line, e.\ g.\
through allowing finite slippage. In this paper, we rigorously
derive a scaling law for the spreading of the droplet in an
intermediate time regime. This scaling law depends only
logarithmically on the length scale introduced by the
regularization, in agreement with a conjecture of de Gennes
\cite{dG}.
\item The rate of coarsening in spinodal decomposition \cite{co}
(with R.\ V.\ Kohn). Spinodal decomposition is usually modelled by
a Cahn-Hilliard equation. In the later stages, it is
experimentally observed that the phase distribution coarsens in a
statistically self-similar fashion. In this paper, we rigorously
prove upper bounds for this coarsening process. The exponents are
the ones heuristically expected and depend on whether the mobility
is degenerate or non-degenerate: $t^{1/4}$ resp. $t^{1/3}$. In
\cite{EO}, we predict a cross-over for almost degenerate mobility
due to a change in the coarsening mechanism.
\item The first-order correction to the Lifshitz-Slyozov-Wagner theory
for Ostwald ripening \cite{HNO} (with A.\ H\"onig and B.\
Niethammer). Ostwald ripening describes the late stage of spinodal
decomposition in an off-critical mixture (volume fraction of one
phase $\phi\ll1$). The minority phase then consists of several
particles immersed in a matrix of the majority phase. The
particles are approximately spherical and don't move---the
Lifshitz--Slyozov---Wagner theory describes the evolution of the
radii distribution. There is a major interest in identifying the
next-order correction term in $\phi$. We rigorously show that
there is a cross-over in the correction term from $\phi^{1/3}$ to
$\phi^{1/2}$ depending on the system size.
\end{itemize}

Our method to rigorously analyze these scaling laws in a
multiscale model is based on relating integral quantities
(energies, average length scales, dissipation rates...). It is
different from the more local method of matched asymptotic
expansions. In particular, it differs from the latter by the
absence of a specific Ansatz. In order to relate the integral
quantities in our {\it Ansatz-free} approach, we need {\it
interpolation inequalities}. These interpolation inequalities
encode the competition of the dominant physical mechanisms in a
scale-invariant fashion (e.\ g.\ the competition between driving
energetics and limiting dissipation or between bulk and surface
energy). Hence tools from pure analysis are here employed in a
more applied context.

In order to illustrate this set of ideas, we present a simple application.

\section*{2. An example from micromagnetics} \addsec

\vskip-5mm \hspace{5mm}

According to the well-accepted micromagnetic model, the
experimentally observed ground-state of the magnetization $m$ is
the minimizer of a variational problem. We are interested in
transition layers (``walls'') between domains in a film of
thickness $t$ in the $(x_1,x_2)$-plane. We assume that the
in-plane axis $m_2$ is favored by the crystalline anisotropy so
that domains of magnetization $m=(0,1,0)$ or $m=(0,-1,0)$ form. In
order to avoid ``magnetic poles'', the walls separating such
domains are parallel to the $x_2$-axis. We are interested in their
specific energy per unit length in $x_2$-direction. Hence the
admissible magnetizations $m$ are $x_2$-independent and connect
the two end-states
\begin{equation}\label{defad}
\begin{array}{c}
m\,=\,m(x_1,x_3)\,\in\,S^2\quad
\mbox{for}\;(x_1,x_3)\,\in\,
\Omega:=(-\infty,\infty)\times(-\frac{t}{2},\frac{t}{2})\\[1ex]
\quad\mbox{and}\quad\lim_{x_1\rightarrow\pm\infty}
m_2(x_1,x_3)\;=\;\pm1.
\end{array}
\end{equation}
The specific energy, which is to be minimized, is given by
\begin{equation}\label{defE}
E(m)\;=\;d^2\int_\Omega|\nabla m|^2\,d^2x
+Q\int_\Omega(m_1^2+m_3^2)\,d^2x
+\int_{\R^2}|\nabla u|^2\,d^2x,
\end{equation}
where $\nabla$ refers to the variables $x=(x_1,x_3)$. Here the
first term is the ``exchange energy'', the second term comes from
crystalline anisotropy and favors the $m_2$-axis. The last term is
the energy of the stray-field $h_s=-\nabla u$ determined by the
static Maxwell equations
$$
\nabla\times h_s\;=\;0\quad\mbox{and}\quad\nabla\cdot(h_s+m)\;=\;0,
$$
which are conveniently expressed in variational form for the
potential $u$
\begin{equation}\label{defuvar}
\int_{\Omega}m\cdot\nabla\zeta\,d^2x
\;=\;\int_{\R^2}\nabla u\cdot\nabla\zeta\,d^2x
\quad\mbox{for all}\;\zeta\in C^\infty_0(\R^2).
\end{equation}
We see that both ``volume charges'' ( $\nabla\cdot m$ in $\Omega$)
and ``surface charges'' ($m_3$ on $\partial\Omega$) generate the
field $h_s$ and thus are penalized. Since the energy density, i.e.
$|\nabla u|^2$, depends on $m$ through (\ref{defuvar}), the
problem is non-local. The constraint of unit length, see
(\ref{defad}), makes the variational problem nonconvex.

The model is already partially non-dimensionalized: The
magnetization $m$ and the field $-\nabla u$ are dimensionless, but
length is still dimensional. In particular, $d$ has dimensions of
length (the ``exchange length'') and $Q$ is dimensionless (the
``quality factor''). Hence the model has two intrinsic length
scales (material parameters), namely $d$ and $d/Q^\frac{1}{2}$,
and one extrinsic length scale (sample geometry), namely $t$.
Despite its simplicity, it is an example of a multiscale model and
we expect different regimes depending on the two nondimensional
parameters $Q$ and $\frac{t}{d}$.

We will focus on the most interesting regime
of ``soft'' materials (i.\ e.\ with low crystalline anisotropy) and
thicknesses $t$ close to the exchange length $d$
\begin{equation}\label{regime}
Q\;\ll\;1\quad\mbox{and}\quad Q\;\ll\;(\frac{t}{d})^2\;\ll\;Q^{-1}.
\end{equation}
Numerical simulation suggest a cross-over {\it within} this range
\cite[Chapter 3.6,Fig.\ 3.81]{HubertSchaefer}:
\begin{itemize}
\item For thin films: ``N\'eel walls''
(see \cite[Chapter 3.6 (C)]{HubertSchaefer}), whose geometry is
asymptotically characterized by
\begin{equation}\label{An1}
\frac{\partial m}{\partial x_3}\,\equiv\,0\;\;\mbox{and}\;\;
m_3\,\equiv\,0
\quad\Longrightarrow\quad m\,=\,(\cos\theta(x_1),\sin\theta(x_1),0).
\end{equation}
\item For thick films:
``Asymmetric Bloch walls'' (see \cite[Chapter 3.6 (D)]{HubertSchaefer}),
whose geometry is asymptotically characterized by
\begin{equation}\label{An2}
\begin{array}{c}
-\nabla u\,\equiv\,0\quad\Longrightarrow\quad
\nabla\cdot m\,=\,0\;\mbox{in}\;\Omega
\;\;\mbox{and}\;\;m_3\,=\,0\;\mbox{on}\;\partial\Omega\\
\Longrightarrow\quad(m_1,m_3)\,=\,(-\frac{\partial\psi}{\partial x_3},
\frac{\partial\psi}{\partial x_1})\;\;\mbox{for a}\;\psi\;\mbox{with}\;
\psi\,=\,0\;\mbox{on}\;\partial\Omega.
\end{array}
\end{equation}
\end{itemize}
This cross-over in the wall geometry is reflected by a cross-over
in the scaling of the specific wall energy $E$. Our proposition
rigorously captures this cross-over in energy.

%%%%%%%%%%%%%%%%%%%%%%%%%%%%%%%%%%%%%%%%%%%%%%%%%%%%%%%%%%%%%%%%%%%%%%%%%%%%%%

\begin{proposition}\label{lower} In the regime (\ref{regime}) we have
\begin{equation}\label{p.7}
\min_{m\;\mbox{{\tiny satisfies}}\;(\ref{defad})} E(m)
\;\sim\;\left\{\begin{array}{lcc}
d^2                             &\mbox{for}&(\frac{t}{d})^2
\;\simgeq\;\ln\frac{1}{Q}\\
t^2\,\frac{1}{\ln\frac{t^2}{Q\,d^2}}&\mbox{for}&(\frac{t}{d})^2
\;\simleq\;\ln\frac{1}{Q}
\end{array}\right\}.
\end{equation}
\end{proposition}

By $\simgeq$, $\simleq$ we mean $\geq$ resp.\ $\leq$ up to a
generic universal constant and $\sim$ stands for both $\simgeq$
and $\simleq$. This scaling qualitatively agrees with the
numerical study of the energy cross-over in the thickness
\footnote{the $x$-axis corresponds to $\frac{t}{d}$, the $y$-axis
to $\frac{E}{d\,t}$, and $Q=0.00025$} given in \cite[Fig
3.79]{HubertSchaefer}.

{\it Upper} bounds are proved by construction. Here we make the
Ansatz (\ref{An1}), resp.\ (\ref{An2}), and let ourselves be
inspired by the physics literature for the details of the
construction. The matching {\it lower} bound in (\ref{p.7}) states
that one cannot beat the Ansatz---at least in terms of energy
scaling---by relaxing the geometry assumptions (\ref{An1}) or
(\ref{An2}). Therefore Proposition \ref{lower} is a validation of
the predicted cross-over in the geometry. We call this type of
analysis {\it Ansatz-free lower bounds}.

%%%%%%%%%%%%%%%%%%%%%%%%%%%%%%%%%%%%%%%%%%%%%%%%%%%%%%%%%%%%%%%%%%%%%%%%%%%%%%%

\section*{3. Proof} \addsec

\vskip-5mm \hspace{5mm}

The upper bound in Proposition \ref{lower} comes
from the following two lemmas. We only sketch their
proof since our main focus is on lower bounds.

\begin{lemma}\label{up1} For $(\frac{t}{d})^2\ll Q^{-1}$ there exists an $m$
of the form (\ref{An2}) with
\begin{equation}\label{up1.1}
E(m)\;\sim\;d^2.
\end{equation}
\end{lemma}

\begin{lemma}\label{up2} For $(\frac{t}{d})^2\gg Q$ there exists an $m$
of the form (\ref{An1}) with
\begin{equation}\label{up2.2}
E(m)\;\sim\;t^2\,\ln^{-1}\frac{t^2}{Q\,d^2}.
\end{equation}
\end{lemma}

For the lower bound we need to estimate the components $m_1$ and
$m_3$ by E. In Lemma \ref{l2} we control $m_3$ by the stray-field
and exchange energy. More precisely, the stray-field energy
penalizes $m_3$ on $\partial\Omega$ in a weak norm. We interpolate
with the $L^2(\Omega)$-control of $\nabla m$ to obtain
$L^2(\Omega)$-control of $m_3$.
In Lemma \ref{l1} we control the vertical average $\overline{m}_1$
of $m_1$ by stray-field, exchange, and anisotropy energy. More
precisely, the penalization of $\nabla\cdot m$ through the
stray-field energy yields a penalization of
$\frac{d\overline{m}_1}{dx_1}$ in a weak norm. We interpolate with
the $L^2(\Omega)$-control of $\nabla m$ (exchange) to obtain an
estimate on the variation of $\overline{m}_1$. We then interpolate
with the $L^2(\Omega)$-control of $m_1$ (anisotropy) to obtain
$L^\infty(\R)$-control of $\overline{m}_1$.

\begin{lemma}\label{l2} We have for any $m$ satisfying (\ref{defad})
\begin{equation}\label{l2.8}
\int_\Omega m_3^2\,d^2x\;\simleq\;\left(1+(\frac{t}{d})^2\right)\,E(m).
\end{equation}
\end{lemma}

\begin{lemma}\label{l1} In the regime $(\frac{t}{d})^2\gg Q$ we have
for any $m$ satisfying (\ref{defad})
\begin{equation}\label{l1.7}
\sup_{x_1\in(-\infty,\infty)}\overline{m}_1^2(x_1)
\;\simleq\;\left(\frac{1}{t^2}\,\ln\frac{t^2}{Q\,d^2}
+\frac{1}{d^2}\right)\,E(m).
\end{equation}
\end{lemma}

%%%%%%%%%%%%%%%%%%%%%%%%%%%%%%%%%%%%%%%%%%%%%%%%%%%%%%%%%%%%%%%%%%%%%%%%%
{\bf Proof of Lemma \ref{up1}}. The construction is due to Hubert
\cite{Hubert}.
%, which itself was motivated
%by the numerical simulation \cite{}.
We nondimensionalize length
by $t$, i.\ e.\ $t=1$. One can construct\footnote{
Indeed, one possible recipe is to start from
$
\psi(x)\;=\;{\textstyle\frac{1}{2}}-|x|
$
and to modify $\psi$ outside of a neighborhood of the curve
$
\gamma\;=\;\left\{\,(\,{\textstyle\frac{1}{2}}
\sqrt{{\textstyle\frac{1}{4}}-x_2^2}\,,\,x_2\,)\,|\,
x_2\,\in\,[{\textstyle-\frac{1}{2},\frac{1}{2}}]\,\right\}.
$
}
a smooth
$\psi\colon\overline{\Omega}\rightarrow\R$ with
$$
|\nabla\psi|^2\;\le\;1\;\;\mbox{in}\;\Omega,\quad
\psi\;=\;0\;\;\mbox{on}\;\partial\Omega\;
\mbox{and for}\;|x_1|\gg 1,
$$
such that there exists a curve $\gamma\subset\overline{\Omega}$
with
$$
\gamma\;\;\mbox{connects}\;\;(0,-{\textstyle\frac{1}{2}})\;\;\mbox{to}\;\;
(0,{\textstyle\frac{1}{2}})\quad\mbox{and}\quad
|\nabla\psi|^2\;=\;1\;\;\mbox{on}\;\;\gamma.
$$
In line with the Ansatz (\ref{An2}), we define $m\colon\Omega\rightarrow S^2$
via
$$
(m_1,m_3)\;=\;
(-\frac{\partial\psi}{\partial x_3},\frac{\partial\psi}{\partial x_1}),\quad
m_2\;=\;\left\{\begin{array}{c}
-\\
+\end{array}\right\}
\,\sqrt{1-|\nabla\psi|^2}\;
\left\{\begin{array}{c}
\mbox{left}\\
\mbox{right}\end{array}\right\}
\mbox{of}\;\gamma.
$$
Only exchange and anisotropy contribute to the energy:
$$
E(m)\;\sim\;d^2+Q,
$$
which turns into (\ref{up1.1}) in the regime under consideration.

%%%%%%%%%%%%%%%%%%%%%%%%%%%%%%%%%%%%%%%%%%%%%%%%%%%%%%%%%%%%%%%%%%%%%%%%%
{\bf Proof of Lemma \ref{up2}}. Making the Ansatz (\ref{An1}), the
energy simplifies to
\begin{eqnarray}\label{up2.1}
&&E(m)\;=\;d^2\,t\int_{-\infty}^\infty|\frac{dm}{dx_1}|^2\,dx_1
+Q\,t\int_{-\infty}^\infty m_1^2\,dx_1+\int_{\R^2}|\nabla u|^2\,d^2x
\\
&&\le\;t^2\left\{\frac{d^2}{t}\int_{-\infty}^\infty
\frac{1}{1-m_1^2}\,(\frac{dm_1}{dx_1})^2\,dx_1
+\frac{Q}{t}\int_{-\infty}^\infty m_1^2\,dx_1+\int_{\R^2}|\nabla U|^2\,d^2x
\right\},\nonumber
\end{eqnarray}
where $U$ is the harmonic extension \footnote{ The inequality
$\int_{\R^2}|\nabla u|^2\,d^2x\le t^2\int_{\R^2}|\nabla
U|^2\,d^2x$ can best be seen by expressing both integrals in terms
of the Fourier transform $\hat m_1(k_1)$ of $m_1(x_1)$.} of $m_1$
from $\{x_3=0\}$ onto $\R^2$. Hence (\ref{up2.1}) holds for any
extension $U$ of $m_1$. We now have to construct $U$ such that its
restriction $m_1$ satisfies $m_1^2(0)=1$ in order to allow for the
sign change of $m_2$. $\int_{\R^2}|\nabla U|^2\,d^2x$ just fails
to control the $L^\infty$-norm of $U$ and thus of $m_1$---the
counterexample involves a logarithm which we also use in this
construction. The logarithm is cut off at the length scales
$\frac{d^2}{t}\ll\frac{t}{Q}$:
$$
U(x)\;=\;\ln^{-1}\frac{Q\,d^2}{t^2}\,
\ln\sqrt{\min\{(\frac{Q\,|x|}{t})^2+(\frac{Q\,d^2}{t^2})^2,1\}}.
$$
An elementary calculation shows (\ref{up2.2}) for $m_1(x_1)=U(x_1,0)$.
A more detailed
analysis of the reduced variational problem (\ref{up2.1}) is
in \cite{Garcia,Melcher}.

%%%%%%%%%%%%%%%%%%%%%%%%%%%%%%%%%%%%%%%%%%%%%%%%%%%%%%%%%%%%%%%%%%%%%%%%%%
{\bf Proof of Lemma \ref{l2}}. We rewrite (\ref{defuvar}) as
\begin{equation}\label{l2.1}
\int_{\Omega} m_3\,\frac{\partial\zeta}{\partial x_3}\,d^2x
\;=\;\int_{\R^2}\frac{\partial u}{\partial x_3}\,
\frac{\partial\zeta}{\partial x_3}\,d^2x
+\int_{\R^2}\frac{\partial u}{\partial x_1}\,
\frac{\partial\zeta}{\partial x_1}\,d^2x
+\int_{\Omega}\frac{\partial m_1}{\partial x_1}\,\zeta\,d^2x
\end{equation}
and choose the test function
$$
\zeta(x_1,x_3)\;=\;\overline{m}_3(x_1)\,\eta(\hat x_3)\quad\mbox{where}\;
x_3\;=\;t\,\hat x_3
$$
and $\eta\in C^\infty_0(\R)$ is chosen such that
$\frac{d \eta}{d\hat x_3}(\hat x_3)=1$ for
$\hat x_3\in(-\frac{1}{2},\frac{1}{2})$
in order to have
$$
\frac{\partial\zeta}{\partial x_3}(x_1,x_3)\;=\;
\frac{1}{t}\,\overline{m}_3(x_1)\,\frac{d\eta}{d\hat x_3}(\hat x_3)\;=\;
\frac{1}{t}\,\overline{m}_3(x_1)\quad\mbox{for}\;x_3\in
(-\frac{t}{2},\frac{t}{2}).
$$
Hence the term on the l.\ h.\ s.\ of (\ref{l2.1}) turns into
\begin{equation}\label{l2.2}
\int_{\Omega}m_3\,\frac{\partial\zeta}{\partial x_3}\,d^2x
\;=\;\int_{-\infty}^\infty\overline{m}_3^2\,dx_1
\end{equation}
and the first term on the r.\ h.\ s.\ of (\ref{l2.1}) is
estimated as follows
\begin{eqnarray}
\left|\int_{\R^2}\frac{\partial u}{\partial x_3}\,
\frac{\partial\zeta}{\partial x_3}\,d^2x\right|
&\simleq&
\left(\int_{\R^2}(\frac{\partial u}{\partial x_3})^2\,d^2x\;
\frac{1}{t}\,\int_{-\infty}^\infty\overline{m}_3^2\,dx_1\right)^\frac{1}{2}
\nonumber\\
&\le&\left(\frac{1}{t}\,E\,
\int_{-\infty}^\infty\overline{m}_3^2\,dx_1\right)^\frac{1}{2}.
\label{l2.3}
\end{eqnarray}
The two remaining terms are also easily dominated:
\begin{eqnarray}
\left|\int_{\R^2}\frac{\partial u}{\partial x_1}\,
\frac{\partial\zeta}{\partial x_1}\,d^2x\right|&\simleq&
\left(\int_{\R^2}(\frac{\partial u}{\partial x_1})^2\,d^2x
\;\,t\,\int_{-\infty}^\infty(\frac{d\overline{m}_3}{d x_1})^2\,dx_1\right)^\frac{1}{2}
\nonumber\\
&\le&\left(\int_{\R^2}(\frac{\partial u}{\partial x_1})^2\,d^2x\,
\int_\Omega(\frac{\partial m_3}{\partial x_1})^2\,d^2x\right)^\frac{1}{2}
\;\le\;\frac{1}{d}\,E,\label{l2.4}\\
\left|\int_{\R^2}\frac{\partial m}{\partial x_1}\,
\zeta\,d^2x\right|&\simleq&
\left(\int_{\R^2}(\frac{\partial m}{\partial x_1})^2\,d^2x\;
t\,\int_{-\infty}^\infty\overline{m}_3^2\,dx_1\right)^\frac{1}{2}
\nonumber\\
&\le&\left(\frac{t}{d^2}\,E\,
\int_{-\infty}^\infty\overline{m}_3^2\,dx_1\right)^\frac{1}{2}.
\label{l2.5}
\end{eqnarray}
Collecting (\ref{l2.2})--(\ref{l2.5}) and using the Cauchy-Schwarz
inequality gives
\begin{equation}\label{l2.6}
\int_{-\infty}^\infty\overline{m}_3^2\,dx_1
\;\simleq\;\left(\frac{1}{t}+\frac{1}{d}+\frac{t}{d^2}\right)\,E
\;\simleq\;\left(\frac{1}{t}+\frac{t}{d^2}\right)\,E.
\end{equation}

On the other hand, we use Poincar\'e inequality in the
$x_3$-direction which we integrate over $x_1\in(-\infty,\infty)$
\begin{equation}\label{l2.7}
\int_\Omega(m_3-\overline{m}_3)^2\,d^2x
\;\simleq\;t^2\,\int_\Omega(\frac{\partial m_3}{\partial x_3})^2 d^2x
\;\leq\;(\frac{t}{d})^2\,E.
\end{equation}
Now (\ref{l2.6}) and (\ref{l2.7}) combine as desired into (\ref{l2.8}).

%%%%%%%%%%%%%%%%%%%%%%%%%%%%%%%%%%%%%%%%%%%%%%%%%%%%%%%%%%%%%%%%%%%%%%%%%%%%%

{\bf Proof of Lemma \ref{l1}}. In the first step we establish for
$0<\rho\ll\ell$ and $0\le\xi_1-\tilde\xi_1\le\ell$
\begin{equation}\label{l1.5}
\left|       \frac{1}{\rho}\int_{\xi_1}^{\xi_1+\rho}\overline{m}_1\,dx_1
-\frac{1}{\rho}\int_{\tilde\xi_1-\rho}^{\tilde\xi_1}\overline{m}_1\,dx_1\right|^2
\;\simleq\;
\left(\frac{1}{t^2}\,\ln\frac{\ell}{\rho}+\frac{1}{\rho\,t}\right)\,E.
\end{equation}
In order to establish (\ref{l1.5}), we construct an appropriate
test function $\zeta$ for (\ref{defuvar}). We first define
$\zeta$ on the strip $\R\times(-\frac{t}{2},\frac{t}{2})$ as piecewise linear
\begin{equation}\label{l1.1}
\zeta(x_1,x_3)\;=\;\left\{\begin{array}{cclcccl}
0                                     &&\xi_1+\rho      &\le&x_1&   &           \\
\frac{1}{\rho}\,(\xi_1-x_1+\rho)      &&\xi_1           &\le&x_1&\le&\xi_1+\rho \\
1                                     &&\tilde\xi_1     &\le&x_1&\le&\xi_1      \\
\frac{1}{\rho}\,(x_1-\tilde\xi_1+\rho)&&\tilde\xi_1-\rho&\le&x_1&\le&\tilde\xi_1\\
0                                     &&                &   &x_1&\le&\tilde\xi_1-\rho
\end{array}\right\}.
\end{equation}
$\zeta$ is just defined such that
\begin{equation}\label{l1.3}
\int_\Omega m\cdot\nabla\zeta\,d^2x\;=\;
-\frac{t}{\rho}\int_{\xi_1}^{\xi_1+\rho}\overline{m}_1\,dx_1
+\frac{t}{\rho}\int_{\tilde\xi_1-\rho}^{\tilde\xi_1}\overline{m}_1\,dx_1.
\end{equation}

For the r.\ h.\ s.\ of (\ref{defuvar}) we have to extend $\zeta$
onto all of $\R^2$: We harmonically extend $\zeta$ on the upper
and lower half-plane $\R\times(\frac{t}{2},+\infty)$ resp.\
$\R\times(-\infty,-\frac{t}{2})$. We claim
\begin{equation}\label{l1.2}
\int_{\R^2}|\nabla\zeta|^2\,d^2x\;\simleq\;\ln\frac{\ell}{\rho}+\frac{t}{\rho}.
\end{equation}
This yields the following estimate of the r.\ h.\ s.\ of (\ref{defuvar})
\begin{eqnarray}\label{l1.4}
\left|\int_{\R^2}\nabla u\cdot\nabla\zeta\,d^2x\right|
&\le&\left(\int_{\R^2}|\nabla u|^2\,d^2x\,
\int_{\R^2}|\nabla\zeta|^2\,d^2x\right)^\frac{1}{2}\nonumber\\
&\le&\left(E\,
\left(\ln\frac{\ell}{\rho}+\frac{t}{\rho}\right)\right)^\frac{1}{2}.
\end{eqnarray}
Obviously (\ref{l1.3}) \& (\ref{l1.4}) yields (\ref{l1.5}).

We now argue in favor of (\ref{l1.2}).
On the strip $\R\times(-\frac{t}{2},\frac{t}{2})$ we have
\begin{equation}\label{l1.a}
\int_{\R\times(-\frac{t}{2},\frac{t}{2})}|\nabla\zeta|^2\,d^2x
\;\stackrel{(\ref{l1.1})}{=}\;
2\,t\,\rho\,(\frac{1}{\rho})^2\;\sim\;\frac{t}{\rho}.
\end{equation}
The Dirichlet integral of the harmonic extension is estimated
in terms of its boundary value as follows
$$
\int_{\R\times(\frac{t}{2},+\infty)}|\nabla\zeta|^2\,d^2x\;\sim\;
\int_0^\infty\frac{1}{x_3^2}\int_{-\infty}^\infty
(\zeta(x_1+x_3,\frac{t}{2})-\zeta(x_1,\frac{t}{2}))^2\,dx_1\,dx_3,
$$
see \cite[Th\'eor\`eme 9.4, Th\'eor\`eme 10.2]{JLL}. Since
$$
\int_{-\infty}^\infty(\zeta(x_1+x_3,\frac{t}{2})-\zeta(x_1,\frac{t}{2}))^2\,dx_1
\;\stackrel{(\ref{l1.1})}{\sim}\;
\left\{\begin{array}{ccccccc}
\ell              &&\ell&\simleq&x_3&&\\
x_3               &&\rho&\simleq&x_3&\simleq&\ell\\
\frac{x_3^2}{\rho}&&    &       &x_3&\simleq&\rho
\end{array}\right\},
$$
this yields
\begin{equation}\label{l1.b}
\int_{\R\times(\frac{t}{2},+\infty)}|\nabla\zeta|^2\,d^2x
\;\simleq\;\ln\frac{\ell}{\rho}.
\end{equation}
Now (\ref{l1.a}) and (\ref{l1.b}) combine into (\ref{l1.2}).

In the second step, we establish for $\ell\gg\frac{d^2}{t}$ and
$0\le\xi_1-\tilde\xi_1\le\ell$
\begin{equation}\label{l1.6}
|\overline{m}_1(\xi_1)-\overline{m}_1(\tilde\xi_1)|^2\;\simleq\;
\left(\frac{1}{t^2}\,\ln\frac{\ell\,t}{d^2}+\frac{1}{d^2}\right)\,E.
\end{equation}
For this, we observe that
\begin{eqnarray*}
\left|\frac{1}{\rho}\int_{\xi_1}^{\xi_1+\rho}\overline{m}_1(x_1)\,dx_1
-\overline{m}_1(\xi_1)\right|^2&\simleq&
\rho\int_{-\infty}^\infty(\frac{d\overline{m}_1}{dx_1})^2\,dx_1\\
&\leq&\frac{\rho}{t}\int_\Omega(\frac{\partial m_1}{\partial x_1})^2\,d^2 x
\;\leq\;\frac{\rho}{d^2\,t}\,E,
\end{eqnarray*}
so that together with (\ref{l1.5}) we obtain
$$
|\overline{m}_1(\xi_1)-\overline{m}_1(\tilde\xi_1)|^2\;\simleq\;
\left(\frac{1}{t^2}\,\ln\frac{\ell}{\rho}+\frac{1}{\rho\,t}
+\frac{\rho}{d^2\,t}\right)\,E.
$$
We now balance the first and last term by choosing
$\rho=\frac{d^2}{t}\simleq\ell$
and so obtain (\ref{l1.6}).

In the last step, we show (\ref{l1.7})
for $\frac{t^2}{Q\,d^2}\gg1$.
For this we observe that
$$
\int_{-\infty}^\infty\overline{m}_1^2\,dx_1\;\le\;
\frac{1}{t}\,\int_\Omega m_1^2\,d^2x\;\le\;
\frac{1}{Q\,t}\,E.
$$
Hence we obtain together with (\ref{l1.6}) for arbitrary
$\xi_1\in(-\infty,\infty)$
\begin{eqnarray*}
\overline{m}_1(\xi_1)^2&\simleq&
\frac{1}{\ell}\int_{\xi_1-\frac{\ell}{2}}^{\xi_1+\frac{\ell}{2}}
\overline{m}_1^2\,dx_1
+\frac{1}{\ell}\int_{\xi_1-\frac{\ell}{2}}^{\xi_1+\frac{\ell}{2}}
(\overline{m}_1(\xi_1)-\overline{m}_1(x_1))^2\,dx_1\\
&\simleq&\left(
\frac{1}{Q\,\ell\,t}+\frac{1}{t^2}\,\ln\frac{\ell\,t}{d^2}+\frac{1}{d^2}
\right)\,E.
\end{eqnarray*}
Choosing $\ell\;=\;\frac{t}{Q}\;\gg\;\frac{d^2}{t}$,
we balance the two first terms
and so obtain (\ref{l1.7}).

%%%%%%%%%%%%%%%%%%%%%%%%%%%%%%%%%%%%%%%%%%%%%%%%%%%%%%%%%%%%%%%%%%%%%%%%%%%%%%%%

{\bf Proof of Proposition \ref{lower}}. It remains to establish
the lower bound. For further reference we remark that by
Poincar\'e's inequality
\begin{equation}\label{p.1}
\int_{(-\frac{t}{2},\frac{t}{2})\times(-\frac{t}{2},\frac{t}{2})}
(m_i-\overline{m}_i(0))^2\,d^2x\;\simleq\;t^2\,
\int_\Omega|\nabla m_i|^2\,d^2x\;\simleq\;(\frac{t}{d})^2\,E.
\end{equation}
According to (\ref{defad}), we have in particular
$\lim_{x_1\rightarrow\pm\infty}\overline{m}_2(x_1)=\pm 1$ and thus
there exists an $\xi_1$ with $\overline{m}_2(\xi_1)=0$. W.\ l.\ o.\ g.\ we
assume $\xi_1=0$ so that
$\overline{m}_2(0)\;=\;0$.
According to (\ref{p.1}) we obtain
\begin{equation}\label{p.2}
\int_{(-\frac{t}{2},\frac{t}{2})\times(-\frac{t}{2},\frac{t}{2})}
m_2^2\,d^2x\;\simleq\;(\frac{t}{d})^2\,E.
\end{equation}
Furthermore, we have according to Lemma \ref{l2}
\begin{equation}\label{p.3}
\int_{(-\frac{t}{2},\frac{t}{2})\times(-\frac{t}{2},\frac{t}{2})}
m_3^2\,d^2x\;\simleq\;\left(1+(\frac{t}{d})^2\right)\,E.
\end{equation}
Since $1-m_1^2=m_2^2+m_3^2$, the estimates (\ref{p.2}) \& (\ref{p.3}) imply
$$
\int_{(-\frac{t}{2},\frac{t}{2})\times(-\frac{t}{2},\frac{t}{2})}
(1-m_1^2)\,d^2x\;\simleq\;\left(1+(\frac{t}{d})^2\right)\,E.
$$
In view of (\ref{p.1}), this localizes to
\begin{equation}\label{p.4}
1-\overline{m}_1(0)^2\;\simleq\;(\frac{1}{t^2}+\frac{1}{d^2})\,E.
\end{equation}
On the other hand, we have by Lemma \ref{l1}
\begin{equation}\label{p.5}
\overline{m}_1(0)^2\;\simleq\;\left(\frac{1}{t^2}\,\ln\frac{t^2}{Q\,d^2}
+\frac{1}{d^2}\right)\,E
\end{equation}
provided $(\frac{t}{d})^2\gg Q$.
Combining (\ref{p.4}) and (\ref{p.5}), we obtain
\begin{equation}\label{p.6}
1\;\simleq\;\left(\frac{1}{t^2}\,\ln\frac{t^2}{Q\,d^2}
+\frac{1}{d^2}+\frac{1}{t^2}\right)\,E
\;\sim\;\left(\frac{1}{t^2}\,\ln\frac{t^2}{Q\,d^2}
+\frac{1}{d^2}\right)\,E.
\end{equation}
Since we have by elementary calculus that
$$
\frac{1}{t^2}\,\ln\frac{t^2}{Q\,d^2}
\;\left\{\begin{array}{c}\simleq\\\simgeq\end{array}\right\}\;
\frac{1}{d^2}\quad\Longleftrightarrow\quad
\ln\frac{1}{Q}
\;\left\{\begin{array}{c}\simleq\\\simgeq\end{array}\right\}\;
(\frac{t}{d})^2,
$$
(\ref{p.6}) is equivalent to the lower bound in (\ref{p.7}).

%%%%%%%%%%%%%%%%%%%%%%%%%%%%%%%%%%%%%%%%%%%%%%%%%%%%%%%%%%%%%%%%%%%%%%%%%%%%

\noindent {\bf Acknowledgments}. The author thanks A.\ DeSimone, Weinan E.\ , R.\ V.\ Kohn, and S.\ M\"uller for
many stimulating discussions on micromagnetics.

%%%%%%%%%%%%%%%%%%%%%%%%%%%%%%%%%%%%%%%%%%%%%%%%%%%%%%%%%%%%%%%%%%%%%%%%%%%%

\label{lastpage}

\end{document}